\documentclass[aps,prb,preprint,showpacs,groupedaddress]{revtex4}  
\usepackage{graphicx}  

\newcommand{\nn}{\nonumber}
\newcommand{\be}{\begin{equation}}
\newcommand{\ee}{\end{equation}}
\newcommand{\bea}{\begin{eqnarray}}
\newcommand{\eea}{\end{eqnarray}}
\newcommand{\h}{Hamiltonian }

\newcommand{\la}{\langle}
\newcommand{\ra}{\rangle}
\newcommand{\half}{\frac{1}{2}\,}

\newcommand{\etal}{{\em{et}}\,{\em{al}} \,}
\renewcommand{\H}{{\ensuremath{\mathcal{H}}}}
\newcommand{\kt}{k_B T}
\newcommand{\dtau}{\epsilon}
\newcommand{\nbx}[1]{ \bar{n}_{#1}}

\begin{document}

\title{Quantum Monte Carlo  Study of a Disordered 2-D Josephson
Junction Array}

\author{W.~A.~Al-Saidi}
\email{Al-Saidi.1@osu.edu}
\author{D.~ Stroud}
\email{Stroud@mps.ohio-state.edu}
\affiliation{Department of Physics,
The Ohio State University, Columbus, Ohio 43210}

\date{\today}

\begin{abstract}

We have studied the superconducting-insulating phase transition in
a disordered two-dimensional Josephson junction array, using
quantum Monte Carlo techniques.  We consider disorder in both the
capacitive energies and in the values of the offset charges. The
calculated phase diagram shows that the lobe structure of the
phase diagram disappears for sufficiently strong disorder in the
offset charge. Our results agree quite well with previous
calculations carried out using a mean-field approximation.

\end{abstract}




\maketitle



\section{Introduction}

A Josephson junction array (JJA) consists of a collection of
superconducting islands connected by Josephson coupling. The
coupling can arise from tunnel junctions through an insulating
layer, or from the proximity effect.  The arrays themselves can be
produced experimentally in a wide range of geometries, and with a
great variety of individual junction parameters.  They can serve
as valuable model systems for studying quantum phase transitions
 \cite{sondhi,lobb} under conditions such that the experimental
parameters can be readily tuned \cite{fazio}. Recently, it has
also been proposed that small groups of Josephson junctions may
serve as quantum bits (qubits) in quantum information technology.
In this case, the quantum logic operations are performed by
manipulating experimental parameters such as gate voltages or
magnetic fields \cite{MSS}.

If the superconducting islands in a Josephson array are
sufficiently large, the array is believed to become
superconducting in two stages.  First, at a temperature near the
bulk transition temperature, each island becomes superconducting.
Secondly, at a lower temperature, and provided that the array is
at least two-dimensional, thermal fluctuations become sufficiently
weak that a global phase coherence can be established throughout
the array \cite{lobb}.

If, however, the islands in the array are small, such phase
coherence may not be established even at temperature $T = 0$. The
reason is that there is a second energy which becomes important in
small grains, namely the charging energy of the grains
\cite{anderson}.  The crucial physics is then determined by the
competition between this charging energy and the Josephson
coupling energy.  If the charging energy is sufficiently large,
then it becomes prohibitively expensive energetically to transfer
Cooper pairs from grain to grain, and the array becomes an
insulator, even though each grain is in its superconducting state.
If, on the other hand, the Josephson dominates, the array becomes
phase coherent at $T = 0$. In this case, Cooper pairs can tunnel
between neighboring grains and the array as a whole will be in the
superconducting phase.  Thus, by tuning the ratio of the charging
and Josephson energies, one can cause the array to undergo a
quantum phase transition between a superconducting and an
insulating state \cite{sondhi,fazio}.

In the limit of very large number of Cooper pairs per grain, the
Josephson junction array with diagonal charging energy is
equivalent to the Bose Hubbard model (BHM), which describes soft
core bosons hopping on a lattice with on-site Coulomb
interactions.  The BHM has previously been extensively studied by
Fisher \etal  \, \cite{fisher}, who constructed its $T = 0$ phase
diagram. In the absence of disorder, the phase diagram shows two
phases: a Mott insulating phase and a superfluid phase. In the
presence of disorder, an additional phase, known as the Bose glass
phase, emerges \cite{scalettar}.
Recently, the phase diagram of the disordered BHM in $2$-dimensions
(2D) has been studied by Lee \etal using QMC \cite{lee}.

The aim of this paper is to construct the phase diagram for the
superconducting-insulating (SI) transition in a particular model for a
JJA, including the effects of disorder.  Such disorder is clearly
unavoidable in most practical systems. The model Hamiltonian we study
has previously been investigated by several authors in ordered and
disordered arrays, using both mean-field theory (MFT) and quantum
Monte Carlo (QMC) techniques. A central goal of our work is to compare
the MFT of Refs. \cite{grignani,mancini} and QMC results in a
disordered array in order to check the accuracy of the MFT.  We shall
show that the MFT is generally satisfactory, even for disordered
systems.

The remainder of this paper is organized as follows.  In the next
section, we present the model Hamiltonian and
describe the QMC algorithm we use.  Our numerical results are
presented in Section III, followed by a brief concluding
discussion in Section IV.

\section{Model Hamiltonian and QMC Algorithm}

Our model of a  JJA Hamiltonian involves two types of degrees of
freedom: the number of excess Cooper pairs ${\hat{n}_i}$ on the
$i${th} grain, and the phase ${\hat{\phi}}_i$ of the
superconducting order parameter on the $i${th} grain.
${\hat{n}_i}$ and ${\hat{\phi}}_i$ are taken to be
quantum-mechanically conjugate variables with commutation
relations $[{\hat{n}_i}, {\hat{\phi}}_j]= -i \delta_{i,j}$.   We
consider the following model Hamiltonian on a square lattice in
2D: \be \H= \half \sum_iU_{ii} ({\hat{n}_i}-\bar{n}_i)^2 - E_J
\sum_{\la i,j \ra} \cos({\hat{\phi}}_i-{\hat{\phi}}_j).
\label{eq:ham} \ee Here $E_J$ is the Josephson coupling strength
between nearest neighbors denoted by $\la i,j \ra$ (assumed to be
the same for all nearest neighbor pairs),
and $U_{ii}$ is the charging energy of the $i${th} grain. We
expect that $U_{ii} = q^2/C_{ii}$, where $C_{ii}$ represents the
capacitance of the $i${th} grain with respect to ground, and $q =
2e$ is the charge of a Cooper pair. In a more general model, the
charging part of the energy would be written
$\frac{1}{2}U_{ij}(\hat{n}_i - \bar{n}_i)(\hat{n}_j -\bar{n}_j)$,
where $U_{ij}$ is an element of the charging energy matrix. In
earlier calculations using MFT, nearest-neighbor and
next-nearest-neighbor terms have been included, but these are
numerically more difficult to include in QMC than the diagonal terms. Finally, the
quantity $\bar{n}_i$ represents the ``offset charge.'' $\bar{n}_i$
is related to the voltage between the $i${th} grain and a common
ground plane.  In an ordered array, $\bar{n}_i$ should be
independent of $i$. We omit any dissipative terms arising from
Ohmic shunts in the junctions, and we assume that {\em amplitude}
fluctuations of the gap on the individual grains can be neglected;
this neglect of amplitude fluctuations should be reasonable if $T \ll T_{c0}$, the single-grain
transition temperature.  Thus tunneling of charge between
grains involves only Cooper pairs, not single electrons.

In our calculations, we include disorder in two terms in the
Hamiltonian: the diagonal charging element $U_{ii}$ and the
offset charge $\bar{n}_i$.   Disorder in $U_{ii}$ may arise
from randomness in the size of the individual grains,
while disorder in $\bar{n}_i$ could arise from random offset
charges near the superconducting grains.  Such disorder is
unavoidable in practical realizations of Josephson arrays.

It is worthwhile to discuss the effects of the offset charges
$\bar{n}_i$ qualitatively.  Such charges can arise in two ways:
(i) from a voltage applied between the array and the substrate,
and (ii) from random charges in the substrate.  In the latter
case, $\bar{n}_i$ is a random variable.  In arrays with only
a few grains, $\bar{n}_i$ can, in principle, be tuned
individually to desired values.
However, such tuning, although possibly still achievable in principle,
may be difficult to attain in practice in
arrays with many junctions.

In ordered arrays, the most important effect of the offset charge is
to reduce the region of the phase diagram where the Mott insulating
phase is stable.  The reason is that, if $\bar{n}_i \neq 0$, charge
fluctuations become energetically less expensive; it is therefore
correspondingly easier to establish phase coherence and to destroy the
Mott insulator.

For the disordered array, we determine the  phase diagram of the
Hamiltonian (\ref{eq:ham}) using QMC techniques 
\cite{wallin,otterlo,roddick}, and compare the results with 
MFT \cite{mancini}. Following the method of Ref.\
\cite{wallin}, we first map the 2D Hamiltonian \h \H \,
[Eq.~(\ref{eq:ham})], onto a classical $3$D Hamiltonian. This is
accomplished by the standard procedure of transforming the
partition function $Z=\mathrm{Tr} e^{-\beta \H}$, where $\beta =
1/k_BT$, into an Euclidean path integral along the imaginary time
axis $\tau$ from $0$ to $\hbar\beta$. To do the transformation,
one breaks up the time integral into $L_\tau$ small time steps,
each of length $\dtau=\hbar\beta/L_\tau$, and uses the identity $
e^{-\beta \H}= e^{-\dtau \H} \ldots e^{-\dtau \H}$.  Next, a
complete set of eigenstates (for example, the eigenstates of the
Cooper pair number operator for the $i${th} grain) are inserted at
each imaginary time step. The next step is to rewrite the
Josephson term using the Villain approximation \cite{villain} as
$e^{ -x \cos\phi} \approx \sum_{m=-\infty}^{\infty}e^{- x
f(x)(\phi-2\pi m)^2/2}$, where $f(x)=\{ 2 x
\ln[I_0(x)/I_1(x)]\}^{-1}$, and $I_n(x)$ is the modified Bessel
function of order $n$. For large values of $x$, $f(x) \approx 1$.

Finally, the partition function $Z$ can be expressed as $Z=
{\mathrm Tr} e^{-S}$, where the action $S$ is defined as
\bea
S= \half \dtau \sum_{i,j, \tau} &U_{i,j}&
(J^{(\tau)}_{i,\tau}-\bar{n}_{i,\tau})
(J^{(\tau)}_{j,\tau}-\bar{n}_{j,\tau}) \nn \\
&+& \frac{1}{2 E_J \dtau f(E_J \dtau)} \sum_{i,\tau, \alpha=x,y}
|J_{i, \tau}^{(\alpha)}|^2 \label{eq:action}. \eea In
Eq.~(\ref{eq:action}), the new degrees of freedom are integer-valued 
current loops ${J}_{i, \tau}^{(\alpha)} ( \alpha=x,y)$, which live on the nodes
of the $3$-dimensional (xy$\tau$) lattice, and satisfy the
continuity equation $\sum_{\alpha=x,y,}\partial_\alpha
J^{(\alpha)}_{i,\tau}+\partial_\tau
J^{(\tau)}_{i,\tau}=0$ at every lattice point. The time
components of the current operators, $J^{(\tau)}_{i,\tau}$,
represent the Cooper pair number operators $n_{i,\tau}$ along
their world lines.
Note that $S$ is a {\em classical action} in $(d+1)$ dimensions
while  the original $\H$ is a quantum Hamiltonian in
$d$ dimensions.

To evaluate $S$, we use
the standard Metropolis algorithm to generate
configurations of
the currents ${\bf J}_{i,\tau}$ at inverse temperature $\beta$.
It is convenient to work in the grand canonical ensemble.
The system size is assumed to be a parallelepiped with dimensions
$(L ,L, L_\tau)$ along the space and imaginary time axes
respectively, with periodic boundary conditions in all three
directions.

We locate the superconducting-insulating (SI) phase transition by
examining the superfluid density $\rho$, which is proportional to
the stiffness of the system against a twist in the phase. As has
been shown previously \cite{wallin},
$\rho$ can be expressed in terms of the
current variables ${\bf J}_{i,\tau}$:
\begin{equation}
\rho=\frac{1}{L^2 L_\tau} \la \la w^{(x)} \ra \ra. \label{eq:sf}
\end{equation}
Here $w^{(x)}= \left|\sum_{i, \tau} J_{i,\tau}^{(x)}\right|^2$
is the so-called winding number in the x-direction,
and $\left \la \left\la \cdots
\right \ra \right \ra$ denotes both a grand canonical and a
disorder average.  If $\rho$ is non-zero in the thermodynamic
limit of a large system, then there is
long-range phase coherence and the system will be in the
superconducting phase.

\section{Results}

We have carried out our simulations for three different
system sizes: $8^3$, $10^3$, and $14^3$, and have usually
averaged over $\sim 100 - 500$ realizations of the disorder.
For system with no disorder we studied system with a size of $20^3$. 
For each disorder realization, the system is equilibrated
by slowly annealing in temperature,
over about $2000 \times L^2L_{\tau}$ passes through the entire
system for each value of $E_J$ considered, followed by
approximately twice as many steps over which the grand canonical
averages are computed.

\begin{figure}[tb]
\includegraphics[width=15cm]{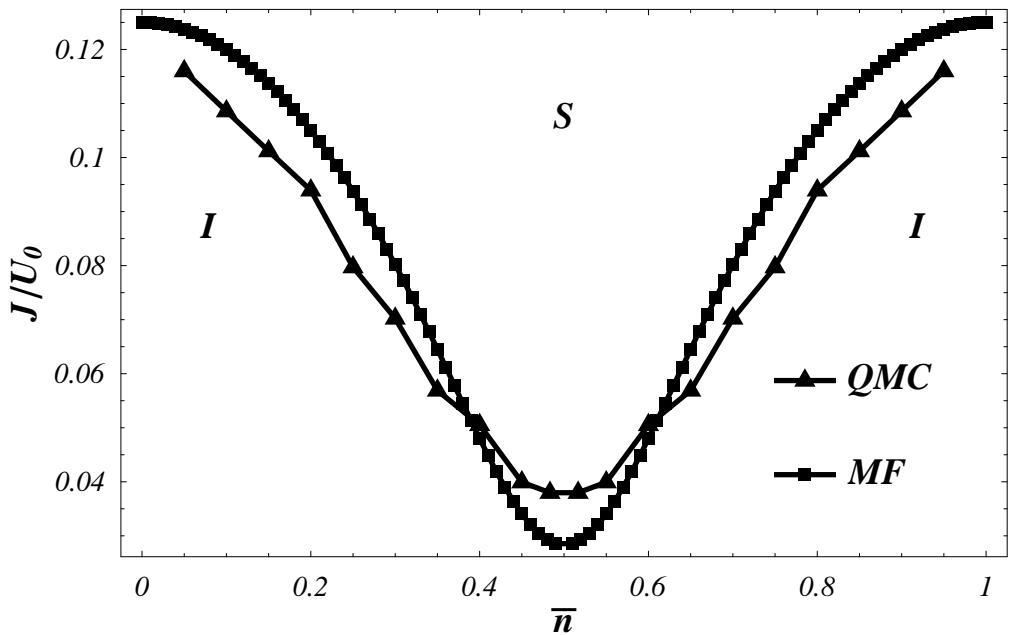}
\caption{ \label{no_dis} Phase diagram for a model
Josephson junction array described by the Hamiltonian (1),
plotted as a function of the offset charge ${\bar{n}}$ as
calculated using QMC and MF theory, in the absence of disorder.
The curves show the critical value of
$J/U_0$ separating the superconducting (S) phase from
the insulating (I) phase.
The temperature is $k_B T = 0.03U $ and the QMC
calculations are done for a lattice of size $20^3$.  The line
segments  simply connect the calculated points}
\end{figure}

Figure \ref{no_dis} shows the phase diagram for a JJA at
$k_B\,T=0.3 U$ with diagonal charging energy and no disorder, as
calculated using QMC as described above.   For reference, we also
show the same phase diagram as calculated using
MFT\cite{grignani,mancini}. The regions of the phase diagram
denoted $S$ or $I$ correspond to the superconducting and Mott
insulating regions, respectively. The two curves show the phase
boundaries as calculated using MFT (squares) and QMC (triangles).
Although in this and later Figures we show results for $0 \leq
\bar{n} \leq 1$, calculations were actually carried out only in
the region $0 \leq \bar{n} \leq 0.5$, since the phase diagram must
be symmetric about $\bar{n}= 0.5$. Our findings here are similar to
what is found by Otterlo \etal in Ref. \cite{otterlo}.

\begin{figure}
\includegraphics[width=15cm]{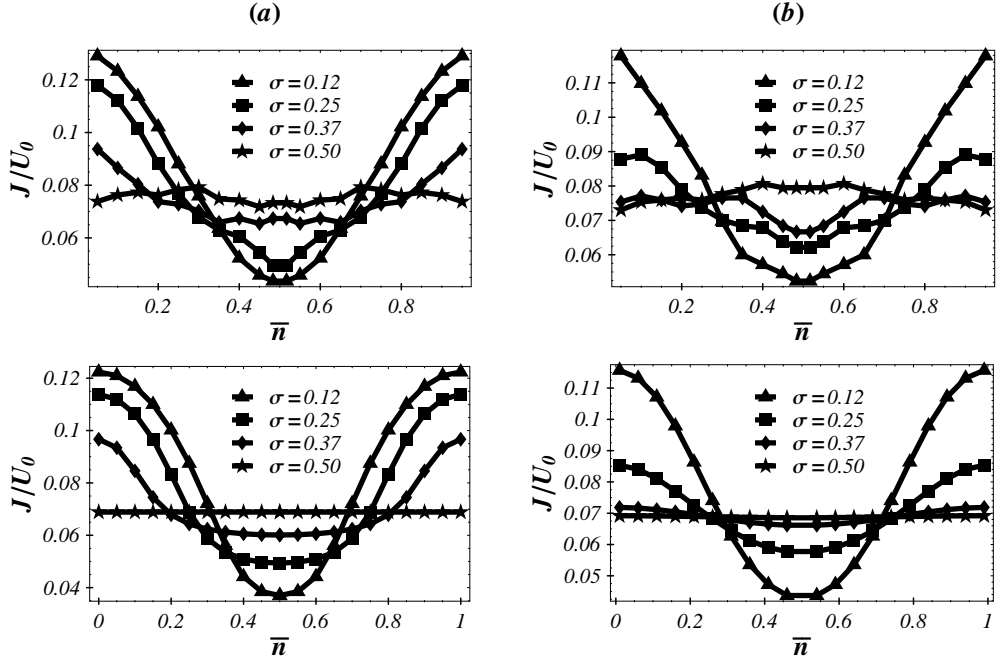}
\caption{\label{Flat_Gauss} (a)
Phase diagram for the model Hamiltonian (1) as calculated using
QMC (upper panel) and MFT theory (lower panel),  including
the effects of disorder.   The offset
charge is chosen from a uniform distribution between
$\bar{n}-\sigma$ and $\bar{n} + \sigma$.
For the QMC calculation, we use a $14^3$ lattice and fix the
temperature at $\kt=0.03 U_0$.  (b) Same as (a) except that the
offset charge is chosen from a Gaussian distribution of
standard deviation $\sigma$ centered at $\bar{n}$.  In these
calculations, line segments connect calculated points.}
\end{figure}

We have considered two types of disorder: randomness in the
offset charges $\bar{n}_i$, and randomness
in the diagonal charging energies $U_{ii}$.
Disorder in $\nbx{i}$ is the analog of chemical potential
disorder in the BHM, while disorder in the $U_{ii}$'s is
analogous to randomness in the mean-field on-site Coulomb
energy in the BHM.

We begin by describing our results for offset charge randomness.
We have carried out calculations with two different random
distributions of $\bar{n}$'s.  In the first case, we choose the
offset charge at each lattice site so that
$\nbx{i}=\bar{n}+\sigma_i$, where $\sigma_i$ is a random number
uniformly distributed in the range $[-\sigma,\sigma]$, where
$\sigma$ represents the strength of disorder. In the second case,
we choose $\bar{n}_i$ to have a Gaussian distribution with a mean
value of $\bar{n}$ and standard deviation $\sigma$, i.e., with a
probability distribution proportional to $e^{-(n_i-{\bar{n}})^2/2
\sigma^2}$.  In both cases, the values of $\bar{n}_i$ on different
sites are taken as uncorrelated.

\begin{figure}
\includegraphics[width=15cm]{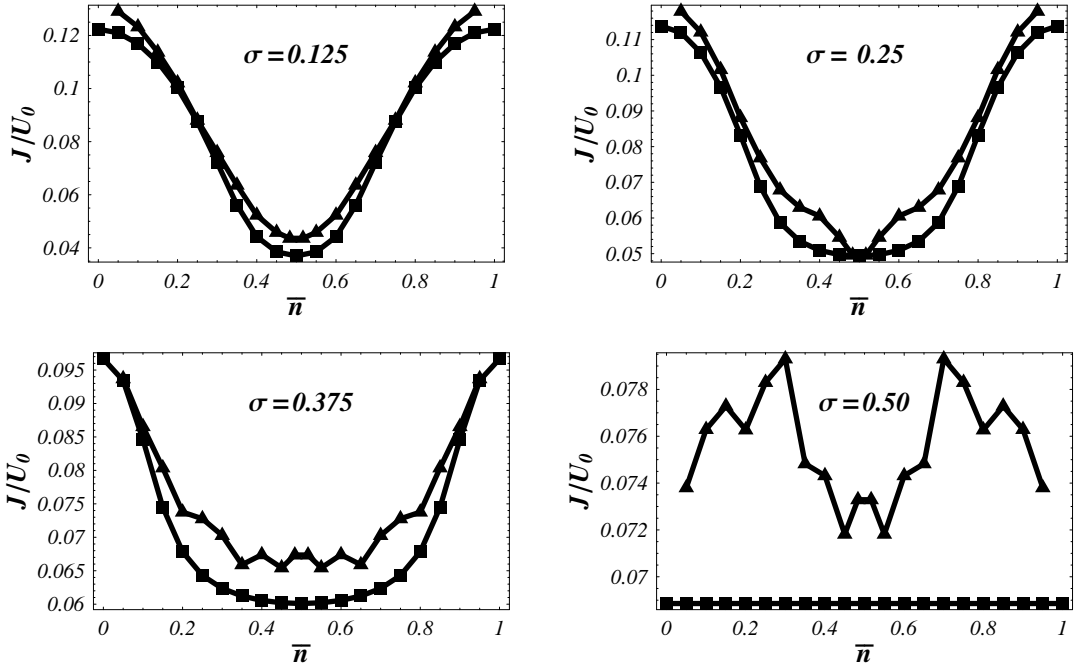}
\caption{\label{comp_flat} Comparison of the phase diagram
as calculated using QMC (triangles) and MF theory (squares),
assuming a uniform distribution of $\bar{n}$ and several values
of the disorder strength parameter $\sigma$, as indicated in
each plot.   Other parameters are
the same as in Fig.~\ref{Flat_Gauss}.}
\end{figure}

\begin{figure}
\includegraphics[width=15cm]{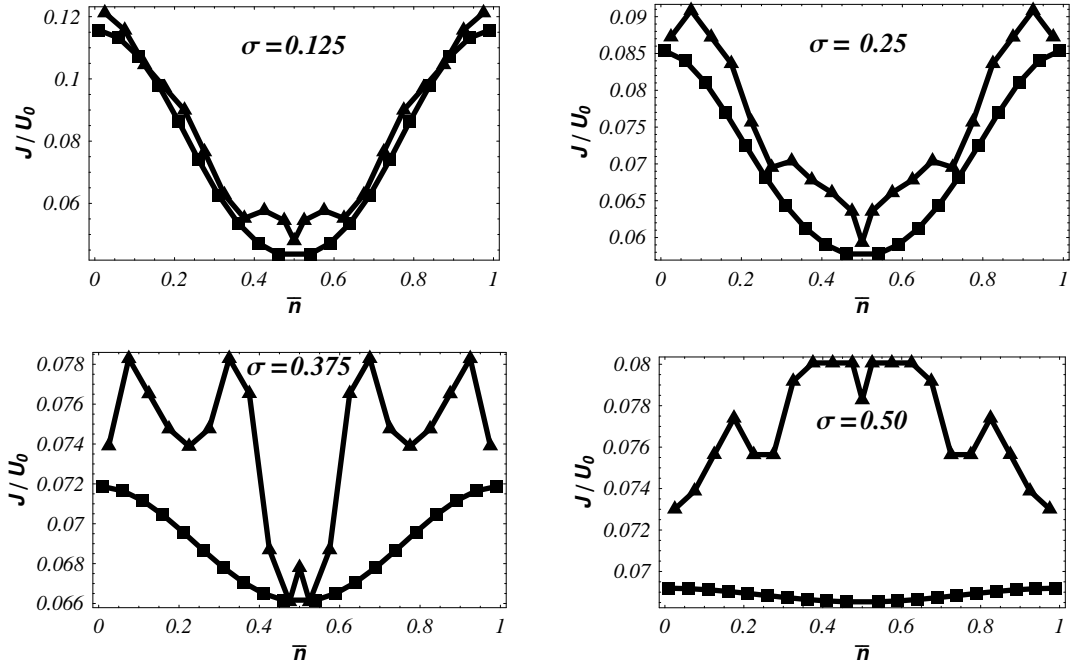}
\caption{\label{comp_gauss} Same as Fig.~\ref{comp_flat} except that
the offset charge ${\bar{n}}$ is chosen from a Gaussian
distribution of standard deviation $\sigma$.}
\end{figure}

\begin{figure}
\includegraphics[width=15cm]{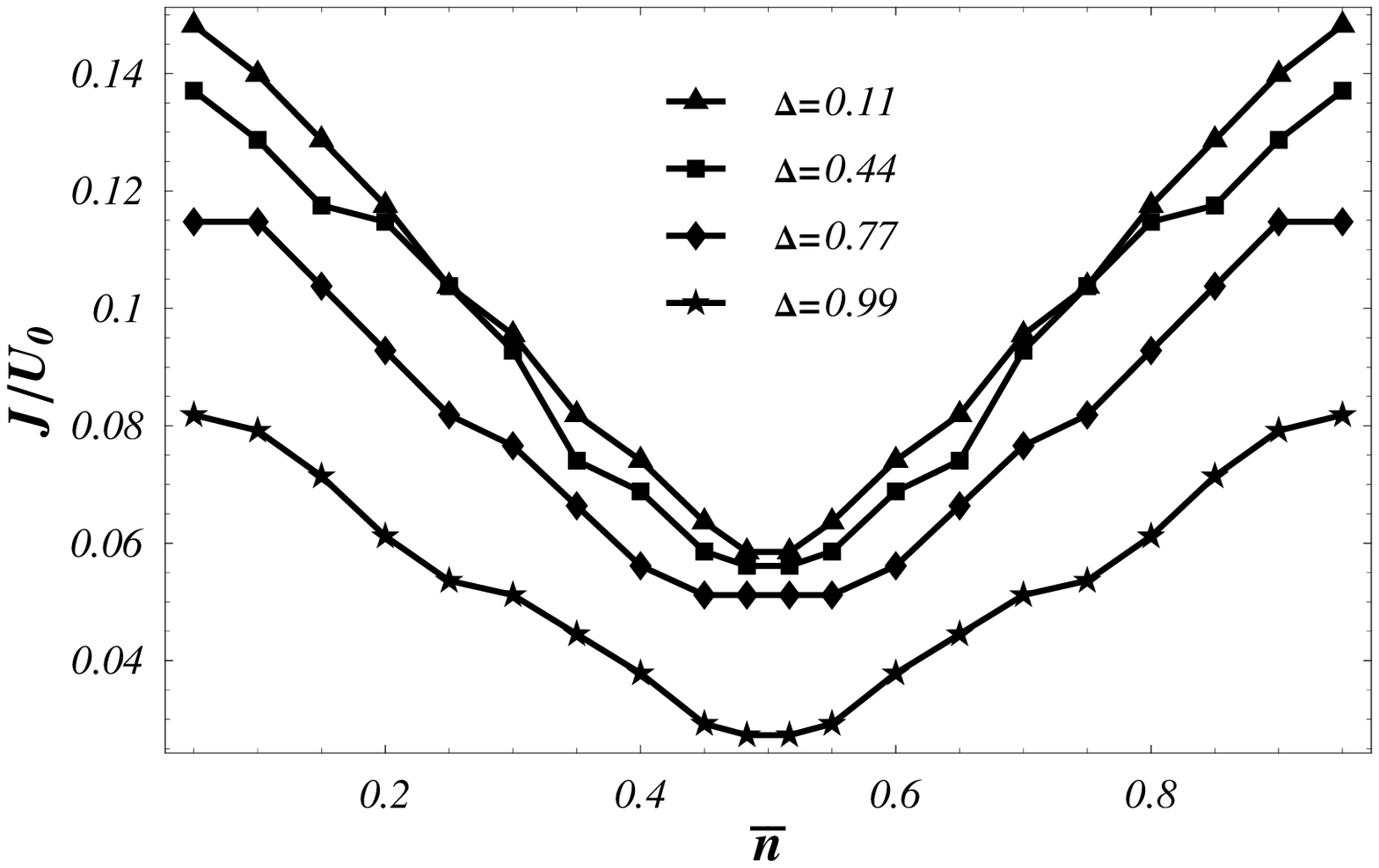}
\caption{\label{dis_u} Phase diagram for a disordered 2D
Josephson junction array described by the Hamiltonian (1),
with charging energy disorder.
Only the diagonal elements $U_{ii}$ of the charging energy
are nonzero; the
$U_{ii}$'s are distributed randomly and uniformly
in the interval [$U_0(1- \Delta)$, $U_0(1+\Delta)]$.
The values of the disorder parameter $\Delta$ are
shown in the Figure.  Other parameters are the same as in
Fig.~\ref{Flat_Gauss}.}
\end{figure}

In Fig.\ \ref{Flat_Gauss}~(a),
we show the phase diagram for a disordered
JJA, assuming a flat distribution of ${\bar{n}_i}$, and as
calculated using QMC (upper panel), and MFT (lower panel).
In each case, the
different curves correspond to the phase
diagrams for different disorder strengths, as
indicated in each Figure.
In Fig.\ \ref{Flat_Gauss}~(b), we show the corresponding
results, but for Gaussian disorder with various standard
deviations $\sigma$, as indicated in the figure.

Several features are noticeable from the QMC calculations. First, at
values of $\bar{n}$ close to $0$ or $1$, disorder reduces the
stability region of the {\em insulating} phase, whereas near $\bar{n}
= 1/2$, disorder {\em increases} the region in which the insulating
phase is stable.  These effects are readily understood: a finite
disorder means that one is effectively calculating the phase diagram
over some average region of $\bar{n}$, and therefore the sharp lobes
seen in the phase diagram for non-random $\bar{n}$ should become less
distinct as the disorder increases.  Indeed, the lobe structure
shrinks and tends to disappear for large values of disorder.  A
similar effect was also emphasized by Fisher \etal for the Bose
Hubbard model \cite{fisher}.  This behavior is not surprising, because
in the limit of large disorder strength $\sigma$, the Hamiltonian
becomes independent of $\bar{n}$: all values of $\bar{n}$ are
therefore equivalent.  (This equivalence is exact for the uniform
distribution at $\sigma = 0.5$.)

Results from the QMC and MFT calculations are compared in a
different way in Figs.\ ~\ref{comp_flat} and ~\ref{comp_gauss} for
the two kinds of disorder in $\bar{n}_i$.
As is evident from both Figures, the MFT results agree very
well with those from QMC results for both types of disorder,
except possibly near the tips of the lobes.  Near
$\bar{n}=0.5$, the critical value of $J/U_0$ is slightly
underestimated by MFT.  A similar underestimate occurs for
large values
of the disorder strength $\sigma$.  In this case, MFT
underestimates the critical value of $J/U_0$ by about
$10$\%.

Figure {\ref{dis_u}} shows the results of calculations
in which disorder is included in the diagonal
charging energy but not in the $\bar{n}_i$'s. 
Specifically, we choose $U_{ii}$ at random from values
uniformly distributed in the interval
$[U_0 (1-\Delta), U_0 (1+\Delta)]$.  We have considered
several values of the disorder strength parameter $\Delta$,
as indicated in the legend.  Clearly, randomness in the
$U_{ii}$ always increases the region of the phase diagram
in which the $S$ phase is stable, whatever the value of $\bar{n}$.
This trend is not surprising, because fluctuations in the
$U_{ii}$'s lead to greater fluctuations in
the number of Cooper pairs on each island.
Thus, by the uncertainty principle, the {\em phase} fluctuations
should be relatively reduced, and therefore the
$S$ (phase-ordered) state should become more favorable,
as seen in our calculations.

\section{Discussion and Summary}

In this paper, we have studied the phase diagram of a model Josephson
junction array at low temperatures using quantum Monte Carlo
techniques.  Our goal was to analyze the effects of disorder in both
the offset charges $\bar{n}_i$ and grain sizes (which affect the
diagonal elements of the charging energy matrix, $U_{ii}$). For
disorder in the $\bar{n}_i$'s, we find that disorder favors the $S$
phase when the average offset charge $\bar{n}$ is close to an integer
number of Cooper pairs, but that it favors the $I$ phase near $\bar{n}
= 1/2$.  We also find that for large disorder, the average value
$\bar{n}$ has little effect on the phase diagram at all.  As for the
second kind of disorder, we find that disorder in the $U_{ii}$'s tends
to favor the $S$ phase for all values of $\bar{n}$, as shown in Fig.\
5.  Finally, we have found that the critical value of $E_J/U$ at the
phase boundary, as calculated from the MFT of Grignani \etal
\cite{grignani}, agrees well with that calculated using QMC (almost
perfectly for small disorder and within $10$\% even for large
disorder). Thus, the MFT gives good results for most of the phase
diagram.

\section{Acknowledgments.}

This work has been supported by NSF Grant DMR01-04987.  The
calculations were carried out on the COE machines of the Ohio
Supercomputer, with the help of a grant of time.


\end{document}